\documentclass{gh}
\usepackage[super,compress]{cite}
\begin{document}

\markboth{Richard Herrmann}{Orthogonal Fractional Cassini Coordinates}

\title{Fractional Cassini Coordinates}
\author{\footnotesize Richard Herrmann}
\address{GigaHedron, Berliner Ring 80, D-63303 Dreieich\\
herrmann@gigahedron.com}

\maketitle

\begin{history}
\received{Day Month Year}
\revised{Day Month Year}
\end{history}

\begin{abstract}
Introducing a set $\{\alpha_i\} \in R$ of fractional exponential powers of focal distances an extension of symmetric Cassini-coordinates on the plane to the asymmetric case is proposed which leads to a new set of fractional generalized Cassini-coordinate systems. Orthogonality and classical limiting cases are derived. An extension to 
cylindrically symmetric systems in $R^3$ is investigated. The resulting asymmetric coordinate systems are well suited to solve
corresponding two- and three center problems in physics.    
\end{abstract}


\ccode{PACS numbers:21.10.Dr;21.60Ev;02.40.Ky}

\tableofcontents

\section{Introduction}
For a given problem an appropriate choice of a specific coordinate system may indeed simplify calculations significantly. 
Special coordinate systems are often used to solve various problems in different areas of e.g. mathematics, natural sciences or engineering. As part of the 
classical canon of standard tools  they are well documented  \cite{moo88}. 

A typical example is the
use of spherical coordinates for isotropic problems or in classical mechanics the use of cylinder coordinates to determine the 
moment of inertia with respect to a given rotational axis. Within atomic physics, in a series of seminal papers Hylleraas \cite{hyl28, hyl29} used prolate elliptic coordinates to calculate the eigenfunctions and values of the two electron Schr\"odinger equation for the helium-atom, applying the Ritz variational principle.

%%%%%%%%%%%%%%%% Figure 3 %%%%%%%%%%%%%%%%%%%%
\begin{figure}[t]
\begin{center}
\includegraphics[width=\textwidth]{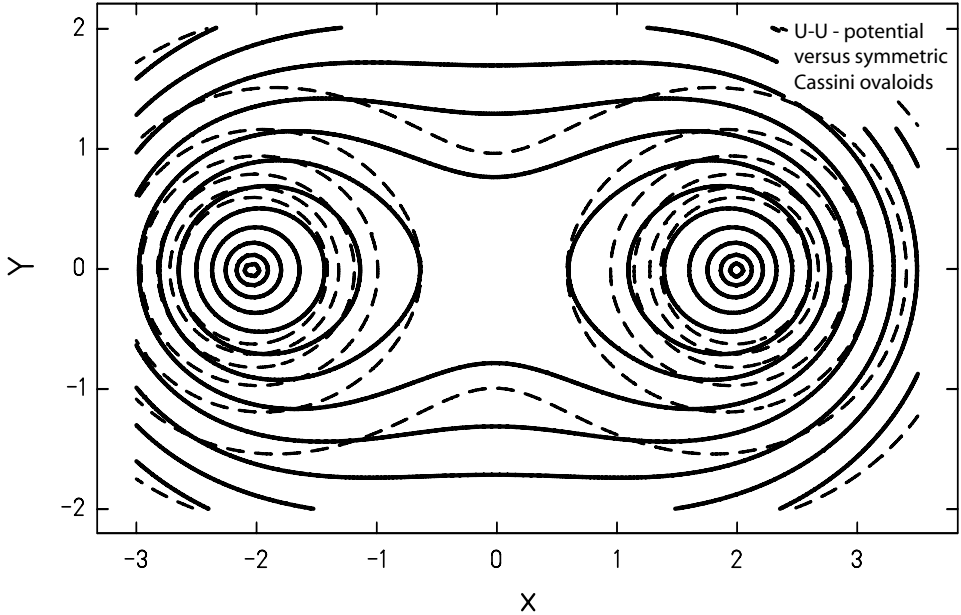}
\caption{
\label{cEKKfig2b}
Contours of the symmetric two-center Coulomb potential (dashed lines) for a fully ionized symmetric U-U quasi-molecule, which, ignoring the finite size and deformation of the atomic nucleus, is given by two point charges $V = -Z_1/r_1 - Z_2/r_2$ with $Z_1=92$ and $Z_2 = 92$, for $-0.4 \leq V  \leq -0.1$ in $\Delta V = 0.033$ steps. Compared to  the iso-$w$ lines
of the symmetric Cassini-coordinate $w$ (solid lines) defined in (\ref{c2250_g1}).
Both families of lines almost coincide. This is largely
independent of the internuclear distance. Hence classical symmetric Cassini coordinates are suitable for both small and large distances between the two nuclei with symmetric charge attributions $Z_1 = Z_2$ \cite{hah17}.  
  }
\end{center}
\end{figure}
%%%%%%%%%%%%%%%%%%%%%%%%%%%%%%%%%%%%%%%%%%%%%%%

While elliptical coordinates are very efficient
in atomic physics for two-center problems for rather small center distances, they become less efficient  with increasing distance between ions\cite{art10}.  For such cases, Cassini-coordinates \cite{cas93} are the better choice, which are defined as:
\index{elliptic coordinates}
\index{Cassini coordinates}
\index{coordinate systems!Cassini}
\index{coordinate systems!elliptic}
\index{coordinate systems!prolate elliptic}
\begin{eqnarray}
\label{c2250_g1}
w    &=& \sqrt{r_1 r_2} \\
\label{c2250_g2}
\theta &=& \frac{1}{2}(\tau_1 + \tau_2 )
\end{eqnarray}
The classical example is
the solution of the two-center Dirac-equation for the study of spectra of quasi-molecules \cite{bet76}, which are formed during slow heavy-ion collisions \cite{slu83, hah17}, since the corresponding potential $V$ is the superposition of two electron-nucleus potentials, see figure \ref{cEKKfig2b}. In this case, $r_i$ is the distance between the electron and the $i$-th nucleus, and
$\tau_i$ is the angle between the inter nuclear axis and the vector $\vec{r}_i$.  
\index{quasi-molecules}
\index{Dirac equation!two center}

In nuclear physics, asymmetric nuclear shapes for large deformation have been calculated using symmetric Cassini ovaloids, where the asymmetry was expanded in a series of Legendre polynomials \cite{pas71, pas08}.
Since symmetric Cassini-coordinates are well suited for symmetric problems, it is tempting and straight forward, to extend the definition of the symmetric Cassini coordinate system to a larger family of  asymmetric orthogonal fractional coordinate systems
in order to describe asymmetric problems appropriately.

\section{Properties of fractional Cassini coordinates}
\index{fractional coordinate systems}
For reasons of simplicity, we first restrict our  presentation of a non symmetric fractional extension of the Cassini-coordinates to $R^2$ and
in a second step  extend the result to rotationally symmetric coordinates in $R^3$. 

Let us assume a collection of  $n$ focal points $F$ in the x-y-plane with coordinates $\{x_i,y_i\}$ and a corresponding set 
of fractional exponents $\{\alpha_i\} \in R$.
We define a new pair of coordinates $x^\mu =  \{w(x,y), \theta(x,y)\}$:
\begin{eqnarray}
\label{c225_g1}
w^{\alpha_s}    &=& \prod_{i=1}^n r_i^{\alpha_i}                      \\
\label{c225_g2}
\alpha_s \theta &=& \sum_{i=1}^n \alpha_i \arctan(\frac{y-y_i}{x-x_i})
\end{eqnarray}
with
\begin{eqnarray}
\alpha_s &=& \sum_{i=1}^n \alpha_i\\
r_i    &=&\sqrt{(x-x_i)^2 + (y-y_i)^2}
\end{eqnarray}
which extend the standard symmetric Cassini coordinate set (\ref{c2250_g1}), (\ref{c2250_g2}) to the fractional case.

The transformation properties of the
$g_{\mu\nu}$ tensor are given as
\begin{eqnarray}
g_{\mu\nu}&=&\frac{\partial{x^i}}{\partial{x^\mu}}
             \frac{\partial{x^j}}{\partial{x^\nu}}
             g_{ij}
\end{eqnarray}

In order to derive the factors
$\frac{\partial{x^i}}{\partial{x^\mu}}$, we apply the derivative operators $\partial_{\mu}$ transformation equations  (\ref{c225_g1}) and (\ref{c225_g2}) and obtain
\begin{eqnarray}
(w^{\alpha_s})_{|\mu} &=& w^{\alpha_s} \{  ( \sum_{i=1}^n \alpha_i \frac{x-x_i}{r_i^2}  ) x_{|\mu}
                      +  ( \sum_{i=1}^n \alpha_i \frac{y-y_i}{r_i^2}  ) y_{|\mu}
                      \}\\
({\alpha_s}\theta)_{|\mu}&=&  \{- ( \sum_{i=1}^n \alpha_i \frac{y-y_i}{r_i^2}  ) x_{|\mu}
                      +  ( \sum_{i=1}^n \alpha_i \frac{x-x_i}{r_i^2}  ) y_{|\mu}
                      \}
\end{eqnarray}
This is a linear system of equations to determine $x_{|\mu}^i$.

With the abbreviations
\begin{eqnarray}
\bar{x}      &=& \sum_{i=1}^n \alpha_i \frac{x-x_i}{r_i^2}   \\
\bar{y}      &=& \sum_{i=1}^n \alpha_i \frac{y-y_i}{r_i^2}
\end{eqnarray}
we obtain in matrix form:
\begin{eqnarray}
\left(
\begin{array}{cc}
w^{\alpha_s} \bar{x}   & w^{\alpha_s} \bar{y}   \\
-\bar{y}   & \bar{x}
\end{array}
\right)
\left(
\begin{array}{c}
x_{|w}  \\
y_{|w}
\end{array}
\right)
&=&
\left(
\begin{array}{c}
{\alpha_s} w^{{\alpha_s}-1}  \\
0
\end{array}
\right)
\\  \nonumber
\\
\left(
\begin{array}{cc}
w^{\alpha_s} \bar{x}   & w^{\alpha_s} \bar{y}   \\
-\bar{y}   & \bar{x}
\end{array}
\right)
\left(
\begin{array}{c}
x_{|\theta} \\
y_{|\theta}
\end{array}
\right)
&=&
\left(
\begin{array}{c}
0  \\
\alpha_s
\end{array}
\right)
\end{eqnarray}
for the derivatives follows:
\begin{eqnarray}
\frac{\partial{x}}
     {\partial{w}}
&=&
\frac{\alpha_s \bar{x}}
     {w(\bar{x}^2 + \bar{y}^2)}  \\
\frac{\partial{y}}
     {\partial{w}}
&=&
\frac{\alpha_s \bar{y}}
     {w(\bar{x}^2 + \bar{y}^2)}  \\
\frac{\partial{x}}
     {\partial{\theta}}
&=&  -
\frac{\alpha_s \bar{y}}
     {(\bar{x}^2 + \bar{y}^2)}  \\
\frac{\partial{y}}
     {\partial{\theta}}
&=&
\frac{\alpha_s \bar{x}}
     {(\bar{x}^2 + \bar{y}^2)}
\end{eqnarray}
and finally for the  $g_{\mu\nu}^{fC}$ tensor for the fractional Cassini coordinates:
\begin{eqnarray}
\label{cassgmnoff}
g_{\mu\nu}^{fC} & = &
\left(
\begin{array}{cc}
\frac{\alpha_s^2}{w^2} \frac{1}{(\bar{x}^2 + \bar{y}^2)}  & 0 \\
0 & \alpha_s^2 \frac{1}{(\bar{x}^2 + \bar{y}^2)}
\end{array}
\right)
\end{eqnarray}
Since the 
$g_{\mu\nu}$ tensor  is diagonal, the coordinate transformation is orthogonal.

\begin{figure}[t]
\begin{center}
\includegraphics[width=\textwidth]{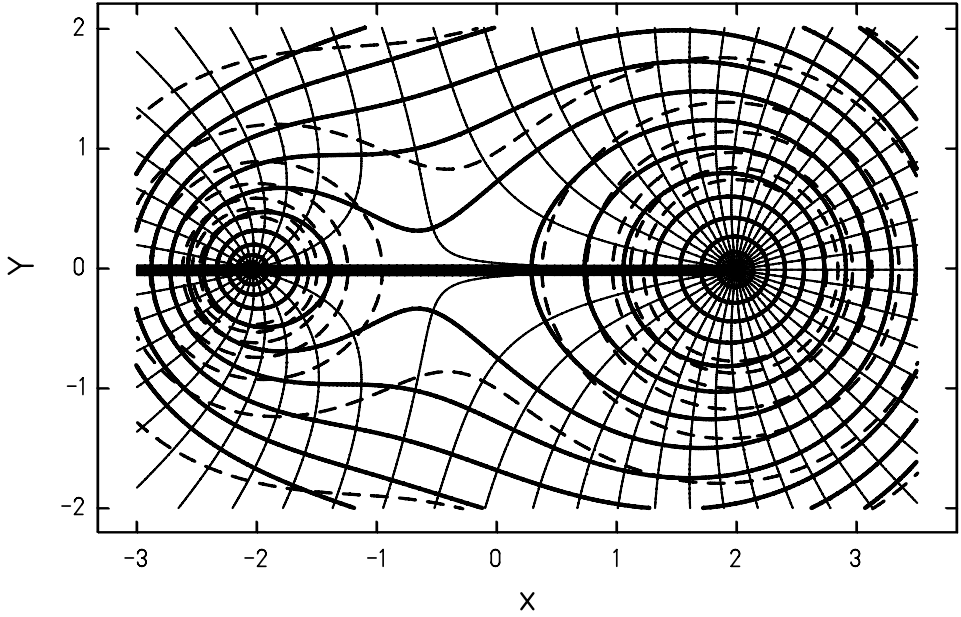}
\caption{
\label{cEKKfig3}
Contours of the orthogonal asymmetric fractional Cassini-coordinate system for two focal points
$F_1 = \{-2,0\}$,
$F_2 = \{+2,0\}$
) and ($\alpha_1 = 0.5$,
$\alpha_2 = 0.92$). 
Solid thick lines show $w = \textrm{const}$, thin lines show $\theta = \textrm{const}$. 
Bold line at $y=0$ indicates the branch cut 
for $\theta = \{0, 2 \pi\}$. Dashed lines show the corresponding two center Coulomb potential with $Z_1 = 50$ and $Z_2=92$.  }
\end{center}
\end{figure}

\begin{figure}[t]
\begin{center}
\includegraphics[width=\textwidth]{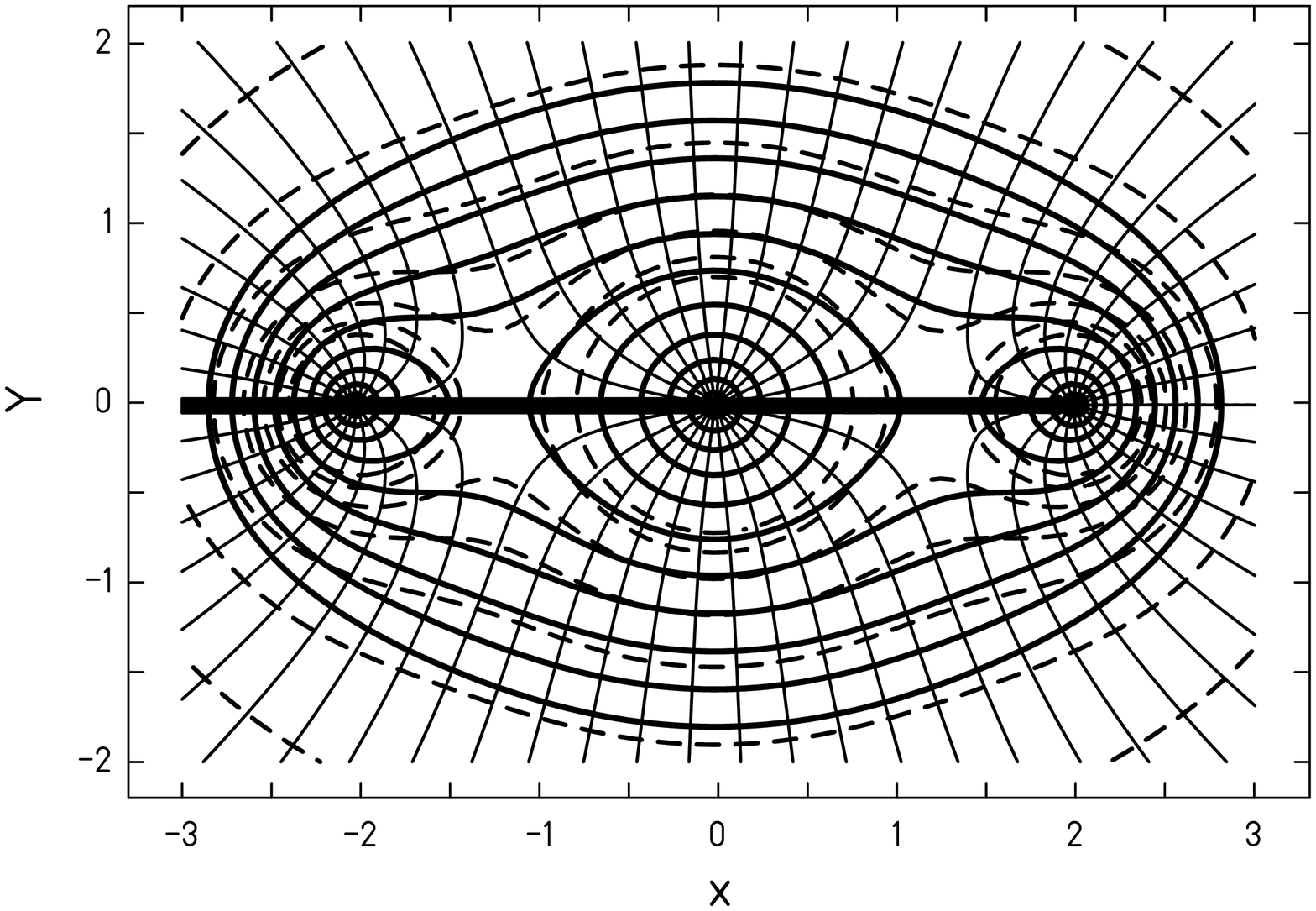}
\caption{
\label{cEKKfig4}
Contours of the orthogonal mirror symmetric fractional Cassini-coordinate system for three focal points (
$F_1 = \{-2,0\}$,
$F_2 = \{0,0\}$,
$F_3 = \{2,0\}$
) and ($\alpha_1 = 3$,
$\alpha_2 = 4$,
$\alpha_3 = 3$). 
Solid thick lines show $w = \textrm{const}$, thin lines show $\theta = \textrm{const}$. 
Bold line at $y=0$ indicates the branch cut 
for $\theta = \{0, 2 \pi\}$.  Dashed lines show the corresponding three center Coulomb potential with $Z_1 = 40$,  $Z_2=82$ and $Z_3 = 40$.}
\end{center}
\end{figure}

\begin{figure}[t]
\begin{center}
\includegraphics[width=\textwidth]{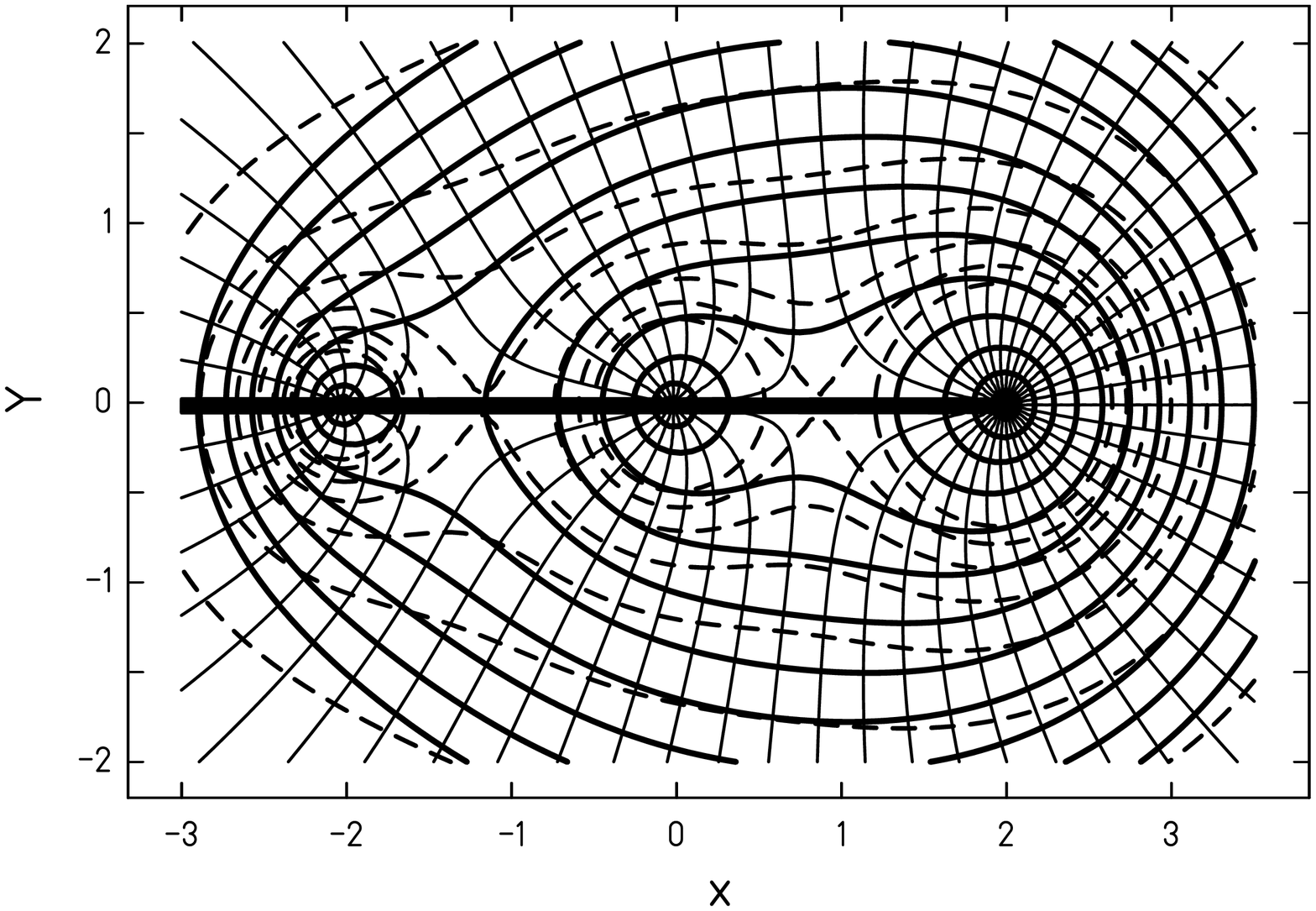}
\caption{
\label{cEKKfig5}
Contours of the orthogonal asymmetric fractional Cassini-coordinate system for three focal points (
$F_1 = \{-2,0\}$,
$F_2 = \{0,0\}$,
$F_3 = \{2,0\}$
) and ($\alpha_1 = 0.2$,
$\alpha_2 = 0.28$,
$\alpha_3 = 0.5$). 
Solid thick lines show $w = \textrm{const}$, thin lines show $\theta = \textrm{const}$. 
Bold line at $y=0$ indicates the branch cut 
for $\theta = \{0, 2 \pi\}$.  Dashed lines show the corresponding three center Coulomb potential with $Z_1 = 20$,  $Z_2=28$ and $Z_3 = 50$.}
\end{center}
\end{figure}

In figures (\ref{cEKKfig3})-(\ref{cEKKfig5}) we compare the orthogonal meshes of constant $\{ w, \theta \}$ (which in nuclear physics may serve as a description of corresponding shapes for a nucleus undergoing a binary or ternary fission process ) to corresponding Coulomb potential $V_c$ for point charges
\begin{equation}
V_c = - \sum_{i=1}^n \frac{Z_i}{r_i} 
\end{equation}
for asymmetric two center, mirror symmetric and asymmetric three center configurations.   
With appropriately chosen values for the powers $\{ \alpha_i\}$  the fractional 
Cassini-coordinates may be adjusted  to follow the equi-potential lines surprisingly well.  
%%%%%%%%%%%%%%%% Figure 3 %%%%%%%%%%%%%%%%%%%%
%%%%%%%%%%%%%%%%%%%%%%%%%%%%%%%%%%%%%%%%%%%%%%%

%%%%%%%%%%%%%%%% Figure 3 %%%%%%%%%%%%%%%%%%%%
%%%%%%%%%%%%%%%%%%%%%%%%%%%%%%%%%%%%%%%%%%%%%%%
%%%%%%%%%%%%%%%% Figure 3 %%%%%%%%%%%%%%%%%%%%
%%%%%%%%%%%%%%%%%%%%%%%%%%%%%%%%%%%%%%%%%%%%%%%
\section{Special cases}
The following special cases with $\{\alpha_i = 1\}$ are included in the definition:
\begin{itemize}
\item[$\bullet$] One focal point $n=1$

polar coordinates with $F_1 = \{0,0\}$
\begin{eqnarray}
\label{cass1Fpol}
g_{\mu\nu}^{pol} & = &
\left(
\begin{array}{cc}
1  & 0 \\
0 & w^2
\end{array}
\right)
\end{eqnarray}
Transformation equations  (\ref{c225_g1}) and (\ref{c225_g2}) simply result as:
\begin{eqnarray}
w    &=& \sqrt{x^2 + y^2}      \\
 \theta &=& \arctan(\frac{y}{x})
\end{eqnarray}

\item[$\bullet$] Two focal points $n=2$

Cassini coordinates with $F_1 = \{-c,0\}$ and $F_2 = \{c,0\}$
\begin{eqnarray}
g_{\mu\nu} & = &
\left(
\begin{array}{cc}
\frac{w^2}{x^2+y^2}  & 0 \\
0 & \frac{w^4}{x^2+y^2}
\end{array}
\right)
\end{eqnarray}

Transformation equations  (\ref{c225_g1}) and (\ref{c225_g2}) follow as:
\begin{eqnarray}
\label{CassSymwt}
w^2    &=& r_1 r_2                  \\
\theta &=& \frac{1}{2} ( \arctan(\frac{y}{x+c}) +
                       \arctan(\frac{y}{x-c}) )
\end{eqnarray}
with
\begin{eqnarray}
r_1    &=&\sqrt{(x+c)^2 + y^2}  \\
r_2    &=&\sqrt{(x-c)^2 + y^2}
\end{eqnarray}

Using the identity (4.4.36) from \cite{Ab}
\begin{eqnarray}
\arctan(u) + \arctan(v) &=&
       \arctan(\frac{u+v}{1-uv})
\end{eqnarray}
we obtain
\begin{eqnarray}
\label{CassSymwtwa}
w^4    &=& (x^2+y^2)^2 -2 c^2 (x^2-y^2) + c^4  \\
\label{CassSymwtwb}
\theta &=& \frac{1}{2} \arctan(\frac{2xy}{x^2-y^2-c^2})
\end{eqnarray}
Now we will determine the term  $x^2+y^2$ in the  $g_{\mu\nu}$ tensor as a function of  $x^\mu$.
In order to simplify procedure we introduce the variables $P$ und $Q$:
\begin{eqnarray}
P   &=& x^2+y^2 \\
Q   &=& x^2-y^2
\end{eqnarray}
Inserting in (\ref{CassSymwtwa}) and (\ref{CassSymwtwb}):
\begin{eqnarray}
w^4    &=& P^2 -2 c^2 Q + c^4  \\
\theta &=& \frac{1}{2} \arctan(\frac{\sqrt{P+Q}\sqrt{P-Q}}{Q-c^2})
\end{eqnarray}
or
\begin{eqnarray}
w^4    &=& P^2-Q^2 + (Q-c^2)^2 \\
\tan^2(2\theta) &=& \frac{P^2-Q^2}{(Q-c^2)^2}
\end{eqnarray}
Explicit we obtain for  $Q$:
\begin{eqnarray}
Q &=& c^2 \pm \frac{w^2}
{\sqrt{1 + \tan^2(2\theta)}}
\end{eqnarray}
and $P$:
\begin{eqnarray}
P &=& \sqrt{w^4+c^4 \pm \frac{2 c^2 w^2}
{\sqrt{1 + \tan^2(2\theta)}}}
\end{eqnarray}
with (4.3.25) and (4.3.26) from \cite{Ab}
\begin{eqnarray}
\frac{1}{\sqrt{1 + \tan^2(2\theta)}} &=& \cos(2\theta)\\
                                    &=& \cos^2(\theta)-\sin^2(\theta)
\end{eqnarray}
$P$ reduces to:
\begin{eqnarray}
\label{CassSymw3}
P &=& \sqrt{w^4+c^4 \pm 2 c^2 w^2 \cos(2\theta)}
\end{eqnarray}
In the limiting case  $c \rightarrow 0$ this  simply should yield polar coordinates. Consequently 
the negative sign in (\ref{CassSymw3}) will be discarded. 

We finally obtain the $g_{\mu\nu}$ for Cassini coordinates
in the standard, familiar form:
\begin{eqnarray}
g_{\mu\nu} & = &
\frac{w^2}{\sqrt{w^4+c^4 + 2 c^2 w^2 \cos(2\theta)}}
\left(
\begin{array}{cc}
1  & 0 \\
0 & w^2
\end{array}
\right)
\end{eqnarray}
\end{itemize}
\section{Extension to cylindrically symmetric  coordinate systems}
Since the explicit form of an orthogonal fractional extension of the Cassini coordinates in two dimensions has been derived,
we finally propose a  transition from $R^2$  (with $g_{\mu\nu}^{(2)}$) to $R^3$ (with $g_{\mu\nu}^{(3)}$), to obtain cylindrically symmetric  fractional coordinate systems,
which are best suited to describe e.g. cylindrically symmetric  two center potentials in atomic physics and asymmetric nuclear shapes  in nuclear fission processes, respectively.

Let us recall, that a coordinate transformation in $R^2$ from Cartesian to polar coordinates, given by
\begin{eqnarray}
x & = & f(r,\theta) \\
\label{CassA3ss}
y & = & g(r,\theta)
\end{eqnarray}
determines the corresponding  two dimensional $g_{\mu\nu}^{(2)}$ tensor:
\begin{eqnarray}
g_{rr}^{(2)} & = & (\frac{\partial f}{\partial r})^2 +
                   (\frac{\partial g}{\partial r})^2 \\
g_{r\theta}^{(2)} & = & \frac{\partial f}{\partial r}
                        \frac{\partial f}{\partial \theta} +
                        \frac{\partial g}{\partial r}
                        \frac{\partial g}{\partial \theta} \\
g_{\theta\theta}^{(2)} & = & (\frac{\partial f}{\partial \theta})^2 +
                   (\frac{\partial g}{\partial \theta})^2
\end{eqnarray}
Rotating this coordinate system around the $x$-axis and introducing cylinder coordinates in $R^3$
$\{\rho,z,\phi\}$,  we apply a  mapping of the above derived $g_{\mu\nu}^{(2)}$ with two dimensional polar coordinates,
where the metric tensor 
$g_{\mu\nu}^{pol}$ is given by (\ref{cass1Fpol}):
\begin{eqnarray}
g_{\mu\nu}^{pol} & = &
\left(
\begin{array}{cc}
1  & 0 \\
0 & \rho^2
\end{array}
\right)
\end{eqnarray}
The transition follows formally by the replacements:
\begin{eqnarray}
x & \rightarrow  & z \\
y & \rightarrow  & \rho 
\end{eqnarray}
It follows
\begin{eqnarray}
\label{CassSym3a}
z & = & f(r,\theta) \\
\label{CassSym3b}
\rho & = & g(r,\theta)
\end{eqnarray}
and
\begin{eqnarray}
\frac{\partial z}{\partial \phi} & = & 0   \\
\frac{\partial \rho}{\partial \phi} & = & 0
\end{eqnarray}
This yields the 
$g_{\mu\nu}$ tensor using the new coordinate set
\begin{eqnarray}
\label{CassSyAA12}
g_{\mu\nu}^{(3)} & = &
\left(
\begin{array}{ccc}
g_{11}^{(2)}  & g_{12}^{(2)} & 0 \\
g_{21}^{(2)}  & g_{22}^{(2)} & 0 \\
0             & 0            & \rho^2
\end{array}
\right)
\end{eqnarray}
The explicit form of $g_{33}^{(3)}$ follows from (\ref{CassSym3b}) and (\ref{CassA3ss}).

Insertion to (\ref{CassSyAA12}) yields the general result for an arbitrarily given $g_{\mu\nu}^{(2)}$:
\begin{eqnarray}
g_{\mu\nu}^{(3)} & = &
\left(
\begin{array}{ccc}
g_{11}^{(2)}  & g_{12}^{(2)} & 0 \\
g_{21}^{(2)}  & g_{22}^{(2)} & 0 \\
0             & 0            & g^{2}(r,\theta)
\end{array}
\right)
\end{eqnarray}
Hence we derived a direct method to transform any two dimensional coordinate system with given $g_{\mu\nu}^{(2)}$ to a three
dimensional cylindrically symmetric coordinate system with $g_{\mu\nu}^{(3)}$.

According to (\ref{cassgmnoff}) off-diagonal elements of the $g_{\mu\nu}^{(2)}$ tensor for the fractional Cassini-coordinates are
vanishing, therefore we obtain finally for asymmetric fractional Cassini-coordinates:
\begin{eqnarray}
g_{\mu\nu}^{fC_3} & = &
\left(
\begin{array}{ccc}
g_{11}^{fC}  & 0 & 0 \\
0  & g_{22}^{fC} & 0 \\
0             & 0            & g^{2}(r,\theta)
\end{array}
\right)
\end{eqnarray}

\section{Conclusion}
We have derived a new family of orthogonal coordinate systems, which extend the symmetric Cassini coordinates
in a reasonable way to the fractional case, introducing a set of fractional exponential coefficients $\{\alpha_i\} \in R$. In addition
we have derived a general procedure to extend orthogonal coordinate systems from $R^2$ to the cylindrically symmetric case
in $R^3$. 

This new set of asymmetric, fractional
coordinate systems may reduce the effort to solve asymmetric two (and more)-center problems in  all branches of physics.

\end{document}